# Toward an Ethics of AI Belief

Authors: Winnie Ma, *Department of Philosophy, King's College London* Vincent Valton, *Institute of Cognitive Neuroscience, University College London*

## 1. Introduction

Philosophical research in AI has hitherto largely focused on the ethics of AI – as well as on other fields such as the philosophy of mind – as opposed to the epistemology of AI. In this paper we, an ethicist of belief and a machine learning scientist, suggest that we need to pursue a novel area of philosophical research in AI – an ethics of belief for AI, i.e., an ethics of AI belief. Here we take the "ethics of belief" to refer to a sub-field within epistemology at the intersection of epistemology and ethics. This subfield is concerned with the study of possible moral, practical, and other non-alethic (non-truth-related) dimensions of belief.

The ethics of belief with respect to human beliefs is an area of philosophical research that has flourished in recent history, as a result of being uniquely placed to address various issues of social justice, particularly around the morality of stereotyping and profiling beliefs in human agents, and discrimination. For example, questions in the ethics of (human) belief on these topics have included: Can agents morally wrong persons just in virtue of holding problematic stereotyping or profiling beliefs about them, apart from/in addition to any prejudicial speech or action we might take? Are agents morally responsible for prejudiced beliefs or implicit biases that they hold? Is it ever morally permissible to hold stereotyping or profiling beliefs, e.g. in the medical domain? Are practical, moral and generally non-alethic considerations relevant with respect to whether we ought to hold particular beliefs, or should our beliefs be based solely on our evidence? What ought we believe when, morally speaking, we ought to hold a particular belief, but practically or evidentially speaking, we ought not, and vice versa, etc.?

Indeed, we came to realize that we need an ethics of AI belief when we realized that a core function of many AI systems is what we've elsewhere called "profiling", or forming beliefs about individuals on the basis of generalizations about a group or groups to which the individual belongs (Ma 2022b). In the case of AI, these will generally be *statistical* generalizations about groups picked out by a particular *feature* (such as age, sex, race, etc.) held in common by all group members.

The infamous COMPAS algorithm, for example, profiles individuals on the basis of generalizations about likelihood of criminal recidivism given features such as family history of imprisonment, number of acquaintances taking illicit drugs, history of fighting in school, etc., prejudicially generating significantly disparate risk scores for White and Black individuals (Angwin et al. 2016). Medical algorithms (as well as human evidence-based medicine practitioners and researchers) also regularly profile individuals on the basis of protected characteristics such as age and sex. Likewise, algorithms profile individuals to determine their risk of default and creditworthiness for loans, mortgages, credit cards, etc.; AI profiles individuals to determine their eligibility for insurance policies, and to set insurance premiums; and so on, algorithms being ubiquitous in a multitude of domains.

Profiling and stereotyping by human agents are belief-practices that are widely discussed in the ethics of belief literature (see, e.g., Rinard 2019; Bolinger 2018; Beeghly 2015; Beeghly forthcoming; Beeghly 2021; Basu 2018; Basu 2019; Basu and Schroeder 2019; Johnson King and Babic 2020). This is in part because they are practices that prima facie generate what have been called "ethical-epistemic dilemmas", in which it seems as though what agents ought to believe from an epistemic point of view conflicts with what they ought to believe from an ethical point of view.[1]

Given that it's clearly at least morally (if not also epistemically) impermissible to profile in a large number of contexts – perhaps most prominently in policing, where racial profiling by police has led to injustices particularly against the Black and Indigenous American communities in the U.S. – we saw the need to query why profiling by AI, while very prevalent, has either not been recognized as this kind of ubiquitous but oftentimes problematic profiling as we've defined the term and therefore in need of further epistemic as well as ethical analysis; or has been assumed to be morally permissible, as it often has been, for example, with respect to medical AI (Ma 2022b; Puddifoot 2019).[2] More generally, our recognition that the epistemic as well as the ethical aspects of such profiling AI required further investigation led us to see that there was a need to pursue research in the novel field of an ethics of AI belief.

In this paper, we will primarily be concerned with the normative question within the ethics of belief regarding what agents – both human and artificial – ought to believe, rather than with descriptive questions concerning whether certain beliefs meet various evaluative standards such as being true, being justified or warranted, constituting knowledge, and so on.[3] We suggest four topics in extant work in the ethics of (human) belief that can be applied to an ethics of AI belief (in §3) *doxastic wronging* by AI beliefs (i.e., AI beliefs that morally wrong the individuals that they concern); morally owed beliefs (beliefs that AI are morally obligated to hold); *pragmatic* and *moral encroachment* on AI beliefs (cases in which the practical or moral features of beliefs are relevant with respect to whether they ought to be held); and moral responsibility for AI beliefs. We also discuss two important, relatively nascent areas of philosophical research that have not yet been recognized as research in the ethics of AI belief, but that do fall within this field of research in virtue of investigating various moral and practical dimensions of belief (in §4): the *epistemic* and *ethical decolonization of AI*; and *epistemic injustice* (injustice involving moral wrongs done to agents in their capacities as knowers or believers) in AI.

## 2. "Belief" in Humans and AI

The first question that needs to be answered to determine whether an ethics of AI belief is possible to begin with is to what extent, if any, AI beliefs may be relevantly analogous to human beliefs in ways that would make an ethics of AI belief possible. Indeed, one of the most philosophically contentious aspects of an ethics of AI belief is the idea that AI have what might be aptly called "beliefs". Other objections to an ethics of AI belief are extensions of objections to any kind of ethics of belief, where this is understood (as we have defined it) as the project of investigating the moral, practical, and other non-alethic dimensions of belief. That is, those who have been called "purists" or "intellectualists" deny that non-alethic considerations are epistemically relevant (see definitions of "purism" and "intellectualism" by, e.g., Grimm 2011, 706; Stanley 2005; Fantl and McGrath 2007; Kim 2017). They thus largely deny that the ethics of belief as we have defined it is a legitimate field of study. A careful rebuttal of such arguments is outside of the current scope of this paper, however.

---

[1] Kelly and Roedder (2008), Gendler (2011), Mugg (2013), and Puddifoot (2017), among others, discuss these so-called "ethical-epistemic dilemmas".

[2] EBM has indeed been *defined* as the "conscientious, explicit, and judicious use of current best evidence [from systematic research] in making decisions about the care of individual patients" (Sackett et al. 1996, 71).

[3] We are here relying on McHugh (2012, 9–10), who draws a similar distinction regarding epistemic norms between *prescriptive* or *deontic norms*, which are norms that have to do with what one ought, may or ought not do; and *evaluative norms*, which have to do with "what is good or bad, valuable or disvaluable", and that only derivatively if at all have to do with what we ought, may, or ought not do.

On the first question, however, that there are such things in AI that have been called "beliefs" by computer scientists is clear: belief is a core concept in many areas of machine learning such as in Probabilistic Graphical Models (Pearl, 1986, 1988; Gelman et al., 2003; Heckerman, 1995; Russell & Norvig, 2003); Deep Belief Networks (Hinton, 2009; Hinton et al., 2006; Y. Bengio et al., 2007; Yoshua Bengio, 2009); Convolutional Deep Belief Networks (Lee, Grosse, et al.; Lee, Largman, et al.); and Restricted Boltzmann Machines (Coates et al., 2011; Hinton, 2010; Hinton & Salakhutdinov 2006; Hinton 1999; Larochelle & Bengio 2008; Smolensky 1986), to name a few. For example, in the study of causal inference, reasoning and learning, Probabilistic Graphical Models are used extensively, where these models are intuitive tools used to encode, reason, and make decisions with uncertain information, and where the term "belief" refers to this "uncertain information" (Benferhat et al., 2020, 2). Elsewhere Pearl (1990, 364) defines "belief" as consisting of "assertions about a specific situation inferred by applying generic knowledge to a set of evidence sentences". Put another way, in these systems, the models' "beliefs" are analogous to a particular conception, which we spell out in more detail below, of human beliefs about complex domains. Probabilistic Graphical Models are furthermore widely used, and include: *belief networks* (a.k.a. Bayes networks, Bayes nets, or decision networks), which are "directed acyclic graphs in which the nodes represent propositions (or variables), the arcs signify direct dependencies between the linked propositions, and the strengths of these dependencies are quantified by conditional probabilities" (Pearl 1986, 241) (see also Lauritzen & Spiegelhalter 1988; Jensen 1996; Darwiche 2009). Some other common Probabilistic Graphical Models include possibilistic networks (Borgelt et al., 2000; Ben Amor & Benferhat, 2005; Benferhat & Smaoui 2007); credal networks (Cozman 2000; Qian et al., 2021); decision trees (de Ville 2013; Kingsford & Salzberg 2008; Raiffa 1968); influence diagrams (Howard & Matheson 2005; Pearl 2005; Shachter 1986); Kappa networks (Halpern 2001); and valuation-based networks (Shenoy 1989; Yaghlane & Mellouli 2008; Xu & Smets 1994). Belief networks can, furthermore, undergo *belief propagation,* where this involves an iterative algorithm for computing marginals of functions on graphical models. Finally, there are also *Bayesian models*, which are probabilistic models that, for example, computational neuroscientists have used to represent optimal human reasoning, or which simply implement Bayesian epistemologies to perform particular tasks. These models can be used in conjunction with experimental testing to measure biases in behavior and estimate the human held beliefs that generate said biases (whether these beliefs are consciously held or not).

Of course, "belief" in the human domain, even in academic fields explicitly focused on belief like epistemology, psychology, etc., is clearly a polysemous term without a fixed or sharp definition.[4] Computer scientist Richard Hadley (1991) likewise notes that the same is true of "belief" in AI.[5] Computer scientist and mathematician Don Perlis, however, says the following about AI "beliefs":

> From a programmer's or logician's point of view, a belief can be thought of simply as a piece of data that has a truth value (it is true or false), and that can in some form be available to an agent to conduct its reasoning, much like an axiom or theorem (Perlis 2000, 362).

Similarly, epistemologist and philosopher of psychology Eric Schwitzgebel offers the following general definition of belief (where this definition appears to have originally been intended for human belief):

---

[4] On the polysemy of "belief" see, e.g., Van Leeuwen & Lombrozo (2023).
[5] Hadley (1991) helpfully catalogues various notions of "belief in AI.

> *belief*: the attitude we have, roughly, whenever we take something to be the case or regard it as true (Schwitzgebel 2021)

For the purposes of this paper, we will follow Perlis and Schwitzgebel in roughly taking a "belief" to be an attitude in which an agent takes some proposition to be true, whether that agent is human or artificial. We will also assume that AI beliefs can be either degreed or probabilistic – i.e., consist of credences – or be full, i.e., binarily true or false, beliefs.

In fact, we realize that oftentimes those working in AI will be more wont to think about beliefs in terms of probability distributions – and at times point estimates – and less wont to think about beliefs in terms of propositions (although seminal figures in artificial intelligence like Pearl (1986) have discussed propositions extensively with respect to AI). Such probability distributions or point estimates will generally be, however, probability distributions representing credences or point estimates representing binary beliefs in particular propositions, such as that "It will rain tomorrow" or that "Patient $X$ has typhus".[6]

More generally, we'll be taking a broadly functionalist and dispositionalist approach to beliefs, holding that what makes something a belief are its actual and potential causal relations to sensory stimulations, behavior, and other mental states.[7] Thus we'll assume that just as what makes a hard drive a hard drive is intuitively not principally a matter of its constitution nor structure (e.g., whether it's a mechanical Hard Disk Drive made of metal platters or a Solid State Drive made of silicon memory chips; or whether it's internally contained or external, etc.), but rather which functions it performs in relation to a computer (e.g., that it stores information, operating systems, applications, etc., that the computer can utilize), so what makes a belief is not principally a matter of its structure but the functions it performs in relation to an agent (whether human or artificial). What the defining functions of beliefs are is, of course, also controversial – see, e.g., Schwitzgebel (2021); Loar (1981); Leitgeb (2014); Zimmerman (2018), etc. – and we will not have the space to adequately argue and specify these definitional functions here. As a first attempt, however, we suggest that a primary causal relationship characteristic of an agent's believing some proposition is for the agent to be disposed to reason and act in ways that are rationally consistent with that proposition on the assumption that the proposition is true. Another defining causal relationship might be that the agent is disposed to associate certain properties with certain entities should those properties and entities be conjoined in sensory input (e.g., forming the perceptual belief that the school bus is yellow, or that the pedestrian is in motion, etc.).

---

[6] In discussion with each other (the co-authors), we have found that answering the question of how we ought to conceptualize belief, or how at least we ought to define "belief" such that our definition would encompass what have been called "beliefs" in the human as well as the AI domains is rather tricky! One co-author says that in machine learning, beliefs are often taken to be probability distributions. The other co-author queries whether it's the case that to think about beliefs as probability distributions is to think merely about the *structure* of beliefs. And this co-author thinks that there is also a need to think about what these probability distributions (which can sometimes be reduced to point estimates) are *about*. This co-author thinks that they are *about* the propositions that she and other philosophers have been used to talk about with respect to beliefs. The other co-author then returns that practitioners working in AI don't tend to think about propositions with truth values. The other co-author wonders, nevertheless, whether the probability distributions constituting beliefs in AI could be helpfully reconceptualized utilizing the notion of propositions. The conversation continues, and both co-authors warmly welcome any feedback and insights that could help clarify these translational issues.

[7] Warmest thanks to those at the 2023 UPenn-Georgetown Digital Ethics Workshop, including and especially Will Fleisher, Elizabeth Edenberg, Gabbrielle Johnson, Lily Hu, Jonathan Flowers, Brian Berkey, Ilaria Cannavotto, and Tae Wan Kim, for pushing us to think more about our notion of "belief". We recognize that what beliefs are – both in humans and AI – has been and remains a contentious issue in the philosophy of mind, epistemology, cognitive science, and in many other fields, and that a lot of fruitful debate about the nature of belief can and should happen at the intersection of philosophy of mind and epistemology / the ethics of belief. We only hope to have made a start at identifying potential functional components common to both humans and AI systems that it would be instrumentally helpful and reasonable to characterize as "beliefs" in the general terms in which we've defined "beliefs". And we recognize that debate about how we ought to define "beliefs" in humans and AI, or whether we ought to resort, rather, to talking about belief-like states in AI, is and should be ongoing.

Having said all of the above in favor of the claim that there is a notion of belief that can be aptly applied to both human and AI beliefs as they've been discussed in their respective domains – a claim that we think is true – we suspect that not much would be lost if we were to instead retreat to saying instead that (current) AI do not have beliefs but rather belief-like states. Perhaps, for example, it's the case that current AI may not be said to have beliefs because they lack the kind of agency required to be aptly termed "believers". As we suggest below, however, perhaps AI will move toward achieving the kind of agency required to be believers as it rapidly evolves. In any case, we suspect that much of what will be said here concerning an ethics of AI belief could instead be said in favor of an ethics of AI "belief-like states" – states that are similar but not identical to genuine beliefs – with marginal loss of significance.

Whether we accept that AI have beliefs or belief-like-states, then – and henceforth, we'll just use the term "belief" to refer to either beliefs or belief-like-states – AI beliefs could include beliefs in belief networks discussed above, where these belief networks may undergo belief propagation. Nodes constituting beliefs in a belief network might look like the following in Figure 1:

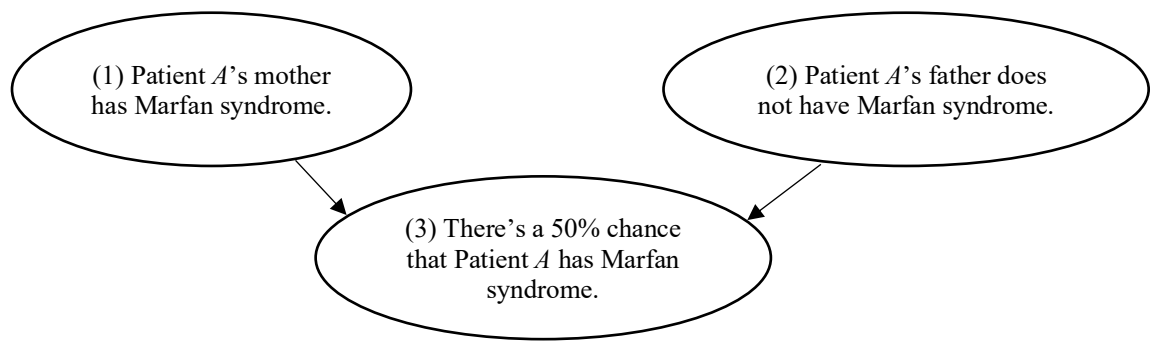

**Figure 1.** This is a simplified representation of a belief network, where nodes (1)-(3) represent beliefs that can take the form of probability distributions or point estimates. We can imagine that belief (3) is arrived at through propagation of beliefs (1) and (2).

AI beliefs could also include perceptual beliefs – for example, beliefs formed by AI in self-driving vehicles about obstacles in the road, road signs, etc.; and information or data, such as that the average house price in San Francisco is $1.2 million; and so on. For the purposes of this paper, we'll be focusing on what we'll call *predictive beliefs* – that is, beliefs of the form "There's a 75% chance that Patient *A* has medical condition *m*", "*B* scores a 5 on the criminal recidivism risk scale", "*C* is highly likely to default on a loan, if it is offered", or "There's a 2% chance that *D* will be involved in a vehicular accident in the next year", and so on. The agents – such as the COMPAS algorithm, or algorithms used by banks to determine eligibility for, say, a credit card or loan– holding such beliefs are indeed generally disposed to reason and act in ways rationally consistent with their belief's propositional content (e.g., by deciding to accept or refuse a credit card application). We take it that these so-called predictive beliefs are both prevalent and one of the most clearly belief-like states that AI possess.

## 3. Proposed Novel Topics in an Ethics of AI Belief

Here we suggest four topics in extant work in the ethics of (human) belief that can be applied to an ethics of AI belief: doxastic wronging by AI; morally owed beliefs; pragmatic and moral encroachment on AI beliefs; and moral responsibility for AI beliefs.

### 3.1. Doxastic Wronging by AI

In the ethics of (human) belief, a growing cohort of philosophers are coming to hold the opinion that human agents can *doxastically wrong* each other – that is, they can morally wrong each other just in virtue of holding certain beliefs about each other. Doxastic wronging is a concept introduced by ethicists of belief Rima Basu and Mark Schroeder (Basu & Schroeder 2019; Basu 2023a; Basu 2023b; Basu 2019a; Basu 2018b). In a series of papers, they suggest that human agents can morally wrong one another not just via their actions or speech, but also in virtue of their beliefs, whether or not these beliefs are acted upon or made known to the person(s) whom the belief is about.

We suggest that it is worth considering that it may also be the case that humans can be morally wronged by beliefs about them formed by artificial agents. (See Figure 2 for a diagram of the various agents and beliefs involved in the doxastic wronging that may be thought to occur in Brisha Borden's case concerning the COMPAS algorithm.) Basu and Schroeder offer various intuition-based arguments in support of their claim that human agents can be doxastically wronged. For example, they argue that a person *A* owes person *B* an apology if and only if *A* has morally wronged *B*. And, intuitively according to Basu and Schroeder, persons who have formed prejudiced or hurtful, etc., beliefs about other persons owe them apologies, regardless of whether the beliefs are made known, expressed, or acted upon. They conclude then that it is possible to doxastically wrong someone, that is, to morally wrong someone in virtue of what they believe about them (Basu and Schroeder 2019, 19).

In fact, we take it that most if not all of the arguments offered by Basu and Schroeder in support of the claim that we can be doxastically wronged ultimately rest on personal intuitions, which we have found, anecdotally, to be divided. For example, one of us finds it very intuitive that Brisha Borden was doxastically wronged in virtue of the predictive profiling belief generated about her in the form of a criminal recidivism risk score by the COMPAS algorithm (Angwin et al. 2016), while one of us does not find this intuitive if the beliefs are never acted upon nor made known. In this case, which demonstrated the racially biased nature of COMPAS's predictions, Brisha, a Black woman whose prior offenses were solely comprised of 4 prior juvenile misdemeanors, was given a risk score of 8; while Vernon Prater, a White man whose prior offenses included 2 armed robberies and 1 attempted armed robbery, was given a risk score of 3 (Angwin et al. 2016).

| The agent morally responsible for the doxastic wronging | The agent that doxastically wrongs / the doxastic wrong-er | The belief that is the *means* through which the doxastic wronging is done | The agent that is doxastically wronged |
|---|---|---|---|
| Judge (?) | Judge | Judge's belief: "Brisha Borden is highly likely to criminally recidivate." | Brisha Borden |
| Developers or deployers of the COMPAS algorithm OR COMPAS algorithm (?) | COMPAS algorithm (?) | COMPAS's predictive belief: "Brisha Borden is highly likely to criminally recidivate." | Brisha Borden |

**Figure 2**. Mapped here are the various agents and beliefs involved in the doxastic wronging that may be thought to occur in Brisha Borden's case.

To elaborate on the scenario, assuming pro-doxastic wronging intuitions, let's say that, on the basis of being made aware of this predictive belief (i.e., the risk score) generated by COMPAS, a judge deciding Brisha's sentencing adopts this belief about Brisha's criminal recidivism risk (and as a result hands down a comparatively longer sentence). In this case, one of us takes it that it is plausible that the judge also doxastically wrongs Brisha in virtue of their belief about her likelihood of recidivism. (Both of us also take it, of course, that Brisha is morally wronged by the unjust sentence handed down.) Furthermore, while the agent who is doxastically wronged in both cases of doxastic wronging by the artificial and the human agent clearly remains the same – Brisha; and it's likewise fairly clear that the *means* by which the doxastic wronging occurs is the *belief* formed by either COMPAS or the judge; it is less clear *who* it is that does the doxastic wronging and *who* is morally responsible for the doxastic wronging in the case of doxastic wronging by the COMPAS algorithm's predictive belief as opposed to the case of the judge's belief.

That is, one might think that it is the developers or the deployers of the COMPAS algorithm that are, ultimately, the agents both perpetrating and morally responsible for the doxastic wrong against Brisha rather than the COMPAS algorithm itself. We might furthermore query whether human agents like the judge can be morally responsible for their beliefs given considerations like doxastic involuntarism (moral responsibility for beliefs will be discussed in greater detail in §3.4 below). But whether the COMPAS algorithm itself is capable of doxastically wronging Brisha and bearing moral responsibility for this doxastic wronging, and more generally whether artificial as opposed to human agents are capable of doxastic wronging and bearing moral responsibility for doxastic wronging are issues that merit further exploration. Both of us co-authors suspect that an agent's capacity to doxastically wrong and bear moral responsibility for doxastic wronging will be dependent on the kind agency they possess. Perhaps when artificial agents become more like human agents in relevant ways, we will be more likely to intuit that they are capable of and morally responsible for doxastic wronging. What one of us takes to be clear, however, is that *some kind* of moral wrong plausibly aptly described as a doxastic wrong – or perhaps as a "shmoxastic wrong" if we take AI to have belief-like states rather than beliefs – has occurred in cases in which problematic predictive profiling beliefs (whether or not they are made known to others or acted upon) are generated by algorithms like COMPAS; and that this sort of moral wrong has not yet been fully recognized.

Furthermore, that Brisha and other individuals unjustly profiled by AI in any domain may experience *doxastic* wronging in addition to action- and speech-related moral wrongings means that there may be an additional dimension of injustice that, we suggest, requires further analysis. Indeed, insofar as we may understand "discrimination" as "the differential treatment based on membership of socially salient groups" (Lippert-Rasmussen 2013, 168), and take a broad view of "treatment", we suggest that doxastic (or "shmoxastic") wrongs perpetrated by "profiling" artificial agents may constitute another important kind of discrimination that likewise merits further investigation and regulation.

### 3.2. Morally Owed Beliefs and Epistemic Behaviors

One way in which various authors have implied that a human agent *A* may morally wrong an agent *B* is by failing to hold beliefs, be in certain doxastic states (where doxastic states include belief, lack of belief or ignorance, and varying levels of credence), or engage in various kinds of epistemic behaviors that *A* morally owes it to *B* to be or engage in (see, e.g., Ferzan 2021; Bolinger 2021). These authors have suggested with respect to the #BelieveWomen movement, for example, that we may morally owe it to women to engage in various kinds of epistemic behaviors in response to women's testimony that they have been sexually assaulted. Ma (2021; manuscript-b) suggests that we may also have moral obligations to be in various doxastic states or engage in certain epistemic behaviors in many other contexts. For example, Ma (2021; manuscript-b) suggests that we can morally owe it to individuals to believe them when they assert their gender identity; we may owe it to ourselves to hold certain positive self-beliefs; we may likewise owe it to certain persons (e.g., our own children) to believe *in* them, particularly when they may not believe in themselves or don't have others who believe in them; we may morally owe it to epistemically marginalized individuals subject to stereotyping and epistemic injustices to be particularly circumspect with respect to the ways in which we assign credences to their testimony (e.g., in she said/he said cases) given the prevalence of implicit bias, etc.; and so on.

Ma (2021, 106–8; manuscript-a) also suggests that just as we can morally wrong persons just in virtue of the beliefs that we hold about them (i.e., doxastically wrong them), so can we morally do-right by persons just in virtue of being in certain doxastic states concerning them – call this *doxastic right-doing*. If we can doxastically do-right by persons, then this may undergird additional moral obligations to hold certain beliefs or be in certain doxastic states. For example, just as one may morally do-right by a neighbor whose house has just burned down by offering them a place to stay while they recover, and morally owe it to them to help in some way, so one may morally do-right by someone asserting their gender identity to believe them and morally owe it to them to be in a doxastic state of belief regarding their assertion.[8]

---

[8] There are numerous cases from the #BelieveWomen movement in which women and other victims of sexual harassment, abuse, and rape responded with thanks to those who posted on social media and communicated elsewhere that they believed them (see, e.g., Ferzan (2021) and Bolinger (2021)). Ma (2021, 106-8) argues that the fact that many of the women who shared their stories of sexual assault responded to these supportive "I believe you's" with thanks and gratitude suggests that those who said "I believe you" morally did-right by these women, that we can owe it to persons to be in certain doxastic states toward them, and hence that we can doxastically do-right by persons when we acquire towards persons the doxastic states that we morally owe it to them to acquire, as well as doxastically wrong them when we fail to do so.

Likewise, we suggest that it may be the case that belief-forming artificial agents may have moral obligations to engage in certain epistemic behaviors or be in certain doxastic states. Such morally owed behaviors and beliefs would likely duplicate many of the kinds of beliefs ostensibly owed by human agents discussed above.

For example, one might think that risk assessment algorithms – such as the Allegheny Family Screening Tool (AFST) discussed by Eubanks (2018), Vaithianathan et al. (2021), Vredenburgh (2023), and elsewhere; the Domestic Abuse, Stalking, and Harassment and Honor-Based Violence (DASH) algorithm (see, e.g., Grogger et al. (2021), Robinson et al., (2016)); and the Vulnerability Abuse Screening Scale (VASS) (Schofield & Mishra, 2003) – used to predict maltreatment of vulnerable individuals such as children, domestic abuse victims, the elderly, respectively, might be under certain moral obligations to treat certain pieces of evidence in particular ways. Such algorithms are often used to help triage social services resources to potential victims of maltreatment based on risk, where testimonial evidence is often used to determine this risk. While such risk assessment algorithms are often problematic given their tendency to utilize predictive profiling techniques that, as Eubanks (2018) and others have pointed out, are subject to bias against already marginalized groups, we think that certain epistemic behaviors or doxastic states may be morally owed by such algorithms in such cases. For example, there may be moral obligations on AI to directly increase credences in the veracity of testimony given by the allegedly victimized child, domestic abuse victim, or elderly person, relative to testimony given by a second party, and decrease credences in contrary testimony given by alleged abusers. Alternatively, there may be moral obligations to change epistemic behaviors toward testimony given by alleged victims versus perpetrators by, for example, decreasing the evidential standards required for belief in the former relative to the latter.[9] These are, of course, contentious issues still being hotly debated in the philosophical literature. Here we don't want to make any claims about what specific epistemic behaviors or doxastic states might be owed – indeed, we take it this will be context-sensitive. We only suggest that in such cases and elsewhere, there may be some epistemic behaviors or doxastic states that may be morally owed.

We might furthermore imagine an algorithm designed to administer online CBT to a patient suffering from low self-esteem. Given that their self-worth is precisely what is doubted by patient, the CBT algorithm may morally owe it to the patient to believe – if only for pragmatic reasons – that they are the kind of entity that should be treated (or interfaced with) in a certain kind of respectful and (at least apparently) benevolent way. Perhaps the algorithm ought also hold the belief that the patient is capable of improving and going on to live a happy and fulfilling life – again if only for pragmatic reasons, even if it's the case that there are no evidential reasons to support this belief. More will be said in the next section on moral and pragmatic encroachment about the sense of "believe" in which artificial agents may be said to be under moral obligations to believe.

To be clear, our intention here is to suggest that AI may morally owe it to agents to be in certain doxastic states or engage in certain epistemic behaviors (such as adjusting credence thresholds, or being particularly cautious of the influence of implicit bias, stereotyping, etc., with respect to our own epistemic behaviors when assigning credences to the testimony epistemically marginalized individuals), just as ethicists of (human) belief have suggested that humans might morally owe it to agents to do so. It is *not* our intention to present and defend a comprehensive set of rules governing the ways in which moral, other non-alethic and alethic considerations influence what we morally owe it to persons to believe or the epistemic behaviors we ought to engage in in all possible cases. Rather, we take it that the circumstances of specific cases (e.g., she said/he said cases) will determine what we morally owe it to the relevant individuals to believe and the epistemic behaviors we ought to engage in; and that these circumstances are often extremely complex.

---

[9] Thanks to an anonymous reviewer for encouraging us to clarify the kinds of epistemic behaviors and/or doxastic states we think AI might be morally obligated to engage or be in – specifically whether we think there are moral obligations on AI to directly change doxastic states or rather the evidential thresholds required for being in certain doxastic states.

We speculate furthermore that, given their less boundedly rational capacities – e.g., their greater processing and memory capacities relative to humans – artificial agents may in fact, perhaps counterintuitively, be subject to *greater* moral obligation to be in various doxastic states. That is, it is debated in epistemology whether humans are subject to any positive obligations to be in certain doxastic states, or whether they are only subject to negative obligations to refrain from holding certain beliefs (Nelson 2010). If there are positive obligations to be in certain doxastic states, however, epistemologists generally agree that our positive obligations to believe, etc., will be constrained by our bounded human rational capacities.[10] We are not expected to be omniscient, for example, by inferring and then coming to hold all beliefs logically entailed by the beliefs that we currently hold. AI, on the other hand, with their relatively greater processing and memory capacities and cognitive proximity to ideally rather than boundedly rational agents, may be under a greater obligation to be in certain doxastic states than humans.[11]

Finally, AI might be thought to owe it to other agents (especially human ones) to hold certain kinds of beliefs governing their actions or their other beliefs and belief-forming processes. That is, with respect to *human* agents, Clifford (1904) famously suggested that we have a *moral* obligation to form beliefs in accordance with our evidence, writing "It is wrong always, everywhere, and for anyone to believe anything on insufficient evidence." The essay in which the maxim is found, "The Ethics of Belief", indeed gave the field of the ethics of belief its name. AI might likewise be morally obligated to form beliefs in the right sorts of ways, perhaps based on evidence as Clifford suggests, or perhaps also based on *non-alethic considerations* – that is, considerations that don't directly indicate that a belief is true, that it constitutes knowledge, or meets some other truth-related norm, such as moral and practical considerations. Other varieties of belief which AI might morally owe it to human to hold could include beliefs about moral and practical norms (like the requirement for self-driving cars to avoid collisions; or perhaps like Asimov's fictional laws of robotics) governing their behavior as well as their beliefs.

### 3.3. Pragmatic and Moral Encroachment on AI Beliefs

Much has been written recently about two controversial, broadly pragmatist views in the ethics of belief: pragmatic encroachment and moral encroachment (see, e.g., Gardiner 2018; Kim 2017; Fritz 2017; Moss 2018; Worsnip 2020; Basu 2019c; Basu 2021; Bolinger 2020; Kvanvig 2011). For our purposes we will understand these two views as follows:

> *pragmatic encroachment*: practical considerations can be relevant with respect to the epistemic status of an agent's beliefs.
>
> *moral encroachment*: moral considerations can be relevant with respect to the epistemic status of an agent's beliefs.

The "epistemic status" of an agent's beliefs may refer to, for example, whether the belief constitutes knowledge, whether it's justified, whether we ought to hold it, etc. Again, in this paper we are focused on the normative epistemic status of whether an algorithm ought to form or output a belief: for example, whether the COMPAS algorithm ought to have formed and output specific risk scores for particular individuals. The rough idea behind one of the more popular varieties of pragmatic and moral encroachment is that the threshold of evidential support required for it to be the case that the agent ought to form or output a belief will depend on the practical or moral stakes of forming the belief, where the greater the stakes, the more evidential support is required.[12] Thus, we suggest that practical considerations (such as potential consequences for sentencing and general practical outlooks for the individual) and moral considerations (such as potential discrimination and doxastic wronging against the individual) should be taken into account in the generation of risk scores output by COMPAS, where pragmatic and moral encroachment may have led to the generation of different, less biased and potentially more accurate predictive beliefs about recidivism risk.

---

[10] See, e.g., Smithies (2015, 2780–1).
[11] Perhaps worth noting that current AI, such as Bayesian belief networks, also often make use of heuristics or shortcuts as boundedly rational agents do, such as considering only a subset of theoretically relevant cases. This is done in order to reduce the complexity of problems being addressed and hence the amount of time and computing power it would take to compute answers to such complex problems.

That is, while recognizing that these encroachment views are controversial with respect to human beliefs,[13] we suggest that AI beliefs (as well as human beliefs) could or should be subject to pragmatic and/or moral encroachment. In other words, we are doxastic impurists with respect to both human and AI beliefs, taking it that non-alethic considerations can and sometimes should be taken into account in the formation of AI beliefs.

For example, in the medical domain, medical algorithms designed to generate probabilistic predictive diagnostic beliefs concerning the likelihood of various medical conditions given a set of symptoms and other inputs (such as the algorithm behind WebMD's Symptom Checker) intuitively should be subject to pragmatic if not also moral encroachment depending on, among other considerations, the severity of various potential diagnoses. Thus, an algorithm designed to generate a predictive belief about whether a patient has a life-threatening form of prostate cancer intuitively should require more or stronger evidence, taking into account the implications of generating false positives versus false negatives, than an algorithm purposed to generate predictions about whether a patient has the flu.[14]

Furthermore, algorithms that generate probabilistic predictive beliefs output predictions where the probability that the prediction is true is specified, usually via a distribution in which 50% is the threshold around which the algorithm takes some proposition to be false (<50%) or true (≥50%): e.g., the algorithm doesn't just predict that Patient *A* has the flu, but that there's a 75% probability that *A* has the flu. In addition, such algorithms often assign a degree of confidence to their probabilistic prediction. As mentioned above (fn. 12), we don't wish to take a stand on how encroachment occurs on AI beliefs (although we favor so-called "threshold" views). But one possible way in which encroachment could occur on such a probabilistic predictive belief would be by adjusting the degree of evidence required for specific degrees of confidence assigned to a believed probabilistic proposition.[15] Thus, given threshold encroachment, a probabilistic predictive belief that there's a 75% probability that Patient *A* has terminal cancer could require a greater number of repeated tests, or more invasive and reliable tests, than the analogous belief that there's a 75% probability that *A* has the flu. See Figure 3 below for more details regarding the probabilistic predictive beliefs, taking the form of distributions, generated by the abovementioned (fictional) algorithm, and the effects of additional testing on the algorithm's confidence in these beliefs.

---

[12] There are other varieties of moral encroachment that do not refer to thresholds but rather, for example, to reasons. For ease of discussion, we focus in this paper on this popular variety of encroachment, which we might call "threshold encroachment". For more on these other varieties of moral encroachment see, e.g., Worsnip (2020). Worsnip (2020), for example, argues that there are at least two ways in which pragmatic encroachment might occur, which he calls "reasons pragmatism" and "threshold pragmatism". According to reasons pragmatism, beliefs are pragmatically encroached upon in that pragmatic considerations make a difference to the epistemic status of beliefs by constituting reasons for or against belief. On the other hand, according to Worsnip's threshold pragmatism, pragmatic considerations determine how much/what threshold amount of evidence is required for justified belief. Alternatively, Gao (2019) argues in support of credal pragmatism, according to which practical considerations affect credences rather than the thresholds on credences. We do not want to take a stand here on which form(s) of encroachment might be occurring, or in other words, how beliefs are being encroached upon; our intention is again just to suggest that some form of moral and pragmatic encroachment may be occurring with respect to AI beliefs.

[13] Philosophers of science have made claims about inductive risk and values in science that are similar to but also importantly different from pragmatic and moral encroachment in the ethics of belief (see, e.g., Elliott and Richards, 2017; Douglas, 2000; and Steel, 2010). The similarity between claims about inductive risk and value-ladenness of science in the philosophy of science and encroachment theses lies in the fact that both take non-alethic considerations to be potentially relevant with respect to the question of what doxastic states we ought to be in. Thanks to Gabbrielle Johnson for pointing out salient differences between encroachment theses in epistemology and arguments from inductive risk in the philosophy of science.

[14] Johnson King and Babic (2021) discuss the need for stake-sensitive decision thresholds in AI.

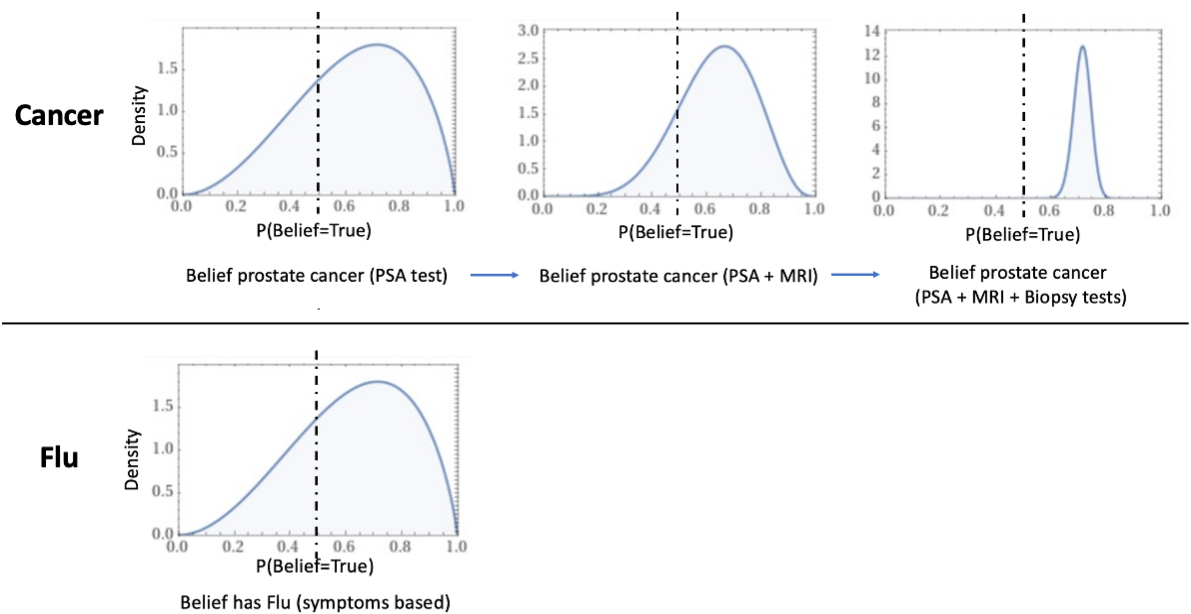

**Figure 3**. These are simplified versions of the kinds of distributions that might be generated by the fictional diagnostic medical algorithm described above designed to generate probabilistic predictive beliefs about the likelihood that a patient has prostate cancer or the flu. 0.5 is the threshold around which the algorithm makes a positive or negative prediction (i.e., "Patient *A* has prostate cancer" if the mode of the distribution falls on or above the 0.5 threshold, or "Patient *A* does not have prostate cancer" if the mode falls below it). The probability assigned to these outright beliefs corresponds to the mode of the distributions. Here for simplicity's sake we've kept the mode at 0.75, although realistically this would also likely shift after repeated tests with differing reliabilities. In the "Flu" case, only one test is done given the generally low stakes of having the flu. And we've stipulated that the test (a simple review of symptoms) is of equal reliability as that of the first test (a PSA test) done for prostate cancer. Given the stipulated equal reliability of the first tests done, the distributions are identical, with each belief being a positive diagnosis assigned a probability of 75%. Then, however, given the greater stakes involved in the prostate cancer case, additional tests (an MRI and then a biopsy) are done. As additional tests are done, the mass of the distribution shifts / the distribution becomes "peakier", indicating greater confidence in the veracity of probabilistic predictive belief.

---

[15] Thanks very much again to an anonymous reviewer for encouraging us to clarify what we took to be appropriate mechanisms via which moral and pragmatic encroachment could occur, and specifically whether we took it to be the case that evidential thresholds ought to be adjusted (as per threshold views) per the moral or practical stakes, or credences/doxastic states directly changed in accordance with the stakes.

Also significant with respect to the practical and moral stakes involved is the intended *use case* of the algorithm. Algorithms purposed to unilaterally make life-or-death decisions in treating patients, for example, will be subject to much higher practical stakes and should therefore have much more stringent evidential requirements than should algorithms that are only meant to be aids to human healthcare workers who will then make decisions concerning patient care; or algorithms like WebMD's Symptom Checker which will have even lower stakes given the effective irrelevance of its outputs to any action that might be taken in official healthcare settings.

We suggested in the last section (§3.2) that artificial agents may be under moral obligations to be in certain doxastic states, whether this means holding certain (degreed or full) beliefs or refraining from holding certain beliefs. We suggest here that one specific way in which it may be that AI morally owe it to human agents to be in certain doxastic states is by forming or holding beliefs that have been appropriately morally encroached upon – that is, beliefs in the formation of which *moral* considerations have been appropriately taken into account. For example, a medical algorithm designed to determine how best to distribute healthcare resources may need to raise the threshold of evidential support required when forming certain beliefs about individuals who are members of racialized and other marginalized groups whose health needs have historically been underserved and whose overall levels of health have generally been comparatively lower than members of dominant groups.

### 3.4. Moral Responsibility for AI Beliefs

Ethicists of belief in the domain of human beliefs are increasingly discussing questions around whether persons may be morally responsible for their beliefs – especially morally problematic racist, sexist, or otherwise bigoted beliefs – and other doxastic states, such as ignorance (see, e.g., Brownstein and Saul 2016; Peels 2017; Mitova 2022; and the <u>Responsible Beliefs: Why Ethics and Epistemology Need Each Other</u> project).

The idea that we can be aptly held morally responsible, morally blamed or praised for the beliefs that we hold or fail to hold is, like many of the other novel claims that we have been discussing, controversial. *Doxastic involuntarism*, or the claim that humans have no or limited voluntary control over what they believe, is often appealed to in conjunction with something like the Ought-Implies-Can principle (according to which being capable of $\varphi$-ing is a necessary condition for a normative obligation to $\varphi$ to apply) in order to defend the view that humans are *not* morally responsible for their doxastic states. That is, the following kind of argument is often presented:

1. For an agent *A* to be morally responsible for being in doxastic state *D*, *A* must be able to voluntarily choose whether or not to be in *D*.
2. Humans are not able to choose whether or not to be in doxastic states due to doxastic involuntarism.

---

3. Therefore, humans are not morally responsible for being in doxastic states.

Human agents' *bounded rationality* – that is, roughly, our limited and flawed reasoning and memory capacities – is also sometimes appealed to in conjunction with the Ought-Implies-Can principle, in order to defend the view that humans are not morally responsible for being in certain doxastic states (Ma 2022b, 171). In particular, Ma (2022b, p.171) and Gendler (2011) argue that because human agents are boundedly rational and yet must generally contend with an overabundance of information and time constraints in making decisions about how to act, we must sometimes make use of certain profiling or stereotyping heuristic beliefs in place of more deliberative System 2-style beliefs.[16]

Note, however, that both excusing considerations – doxastic involuntarism and bounded rationality – do not apply as straightforwardly to artificial as opposed to human agents. That is, Ma (2022b) argues that since AI is at least in principle less boundedly rational than are human agents – i.e., AI process information at greater speeds, have larger memory capacities, and are either less or not subject to many kinds of reasoning errors (e.g., base rate fallacy, confirmation bias, etc.) than are human agents – AI may in fact be *more* morally responsible for their beliefs than human agents.[17] (Future) AI might in fact be helpfully thought of as incarnations of *Homo economicus* or "Econs," the so-called ideally rational agents who do not suffer from limited processing or memory capacities, with whom boundedly rational *Homo sapiens* are contrasted.[18] That is, if moral responsibility inversely correlates with the boundedness of an agent's rationality, AI may (perhaps counterintuitively) bear greater moral responsibility for their doxastic states than do their more boundedly rational human counterparts.[19]

With respect to doxastic voluntarism too, AI, or at least possibly AI of the future, might have a greater degree of voluntary control over their doxastic states. Think, for example, of artificial agents that are able to delete information stored in memory – something that human agents are not able to choose to do directly or voluntarily. There is, of course, the issue currently of whether present-day AI can be aptly said to have *any* voluntary control as their "agency" does not presently amount to the relevant kind of agency and voluntary control over actions that human agents possess. As AI approaches the human ideal of agency and relevant degrees of voluntary control over their beliefs and actions – and perhaps some unsupervised learning algorithms are drawing closer to this ideal – AI might be said to be more doxastically voluntarist than human agents. In this case, then, AI might also be said to be more morally responsible for their beliefs than their more doxastically involuntarist human counterparts, again if moral responsibility for beliefs inversely correlates with degree of doxastic involuntarism.

At the moment, of course, given current distances between the degree of moral-responsibility-conferring agency possessed by present-day AI and humans, the designers or deployers of AI will generally be the ones aptly held morally responsible for the AI beliefs and actions, such as the potentially discriminatory beliefs generated by COMPAS. This may change in the future, however, as AI becomes more powerful in its processing and memory capacities, as well as potentially more autonomous and relevantly "agential". AI models that, once deployed, learn from their interactions and improve themselves without human supervision, for example, might be thought to be moving toward a greater degree of this kind of agency. In the meantime too, as Ma (2022b, 171) suggests with respect to medical AI specifically, less boundedly rational and potentially less doxastically involuntarist AI, with its greater processing and memory capacities and lower risk of falling prey to certain reasoning fallacies, can be and has been used as a complementary tool to aid humans form more accurate beliefs more quickly and consequently also make better decisions as to how to act. Think, for example, of ER doctors making use of medical algorithms as complementary aids to form more accurate beliefs and make better decisions more speedily under severe time constraints and with very high stakes.

---

[16] See, e.g., Kahneman (2012) on System 1 and System 2 type thinking.

[17] We also speculate about whether AI might be *ideally* as opposed to *boundedly emotional* – a notion Ma (2022a) defines as being limited in one's emotional capabilities, notably in the intensity, duration, and possible combinations of one's emotional states; what it would mean to be ideally emotional (e.g., would this mean emotion-less, or rather, capable of feeling any intensity and/or range of emotion for any duration); and what the implications of this would be for the ethics of AI belief, including moral responsibility for AI beliefs. Ma (2022a), for example, argues that the bounded emotionality of human agents can have implications for the doxastic norms that govern them, and in particular can mean that non-alethic considerations can be relevant with respect to what human agents ought to believe.

[18] See, e.g., Thaler (2015) for a discussion of ideally rational *Homo economicus*.

[19] Also exciting to us is the speculative possibility that because AI is in theory ideally rational, artificial agents could be used as tools to complement and aid humans in their moral reasoning. That is, AI might make quite good moral epistemologists, or at least be helpful to humans engaged in moral epistemology!

## 4. Nascent Extant Work that Falls Within the Ethics of AI Belief

Here we discuss two relatively nascent and important areas of philosophical research that haven't yet been generally recognized as ethics of AI belief research, but that do fall within this field of research in virtue of investigating various moral and practical dimensions of belief: the epistemic and ethical decolonization of AI; and epistemic injustice in AI.

### 4.1. Epistemic and Ethical Decolonization of AI

Calls to decolonize AI have risen from all over the world, where "decolonizing" is another polysemous term that is understood differently in different contexts. Here we can understand it roughly as the endeavor to eliminate relationships of domination and subjugation by colonial and colonized communities, respectively. Philosophical work on decolonizing AI has already begun (see, e.g., Mhlambi 2020; Cave and Dihal 2020; Cave 2020; Krishnan et al.; Digital Futures Lab and Research ICT Africa 2022; Research ICT Africa 2022). Much of this philosophical work falls within the purview of the ethics of AI belief in that it deals with non-alethic – e.g., moral, political, practical – dimensions of colonial epistemic structures within which AI may operate or which AI may itself embody.

Our contribution to the growing literature on the decolonization of AI here is very speculative. But, based on other work by Ma (2021; manuscript), we just want to here very briefly suggest a distinction between the *ethical* and *epistemic decolonization* of AI,[20] and suggest ways in which ethical and epistemic decolonization of AI might be pursued. We'll mean the following by the "ethical" and "epistemic decolonization of AI":

> *ethical decolonization of AI*: eliminating relationships of domination and subjugation by colonial and colonized communities' ethical frameworks and norms, respectively, with respect to AI

> *epistemic decolonization of AI*: eliminating relationships of domination and subjugation by colonial and colonized communities' epistemic norms, respectively, with respect to AI

Colonial ethical and epistemic frameworks and norms have dominated the landscape in the development and deployment of AI.

For example, there has recently been a lot of hype about large language models (LLMs) like ChatGPT. Due to the gargantuan degree of resources – including financial resources as well as computing power – required to develop the high-performing LLMs we have now, only entities like large tech companies (such as OpenAI, Microsoft, Google, Facebook, Huawei, etc.) or nations with large economies willing to devote enormous resources to AI research have been able to develop these LLMs (see, e.g., Strubell, Ganesh, and McCallum 2019; Schwartz et al. 2020; and Nikolaiev 2023). Furthermore, LLMs require an enormous amount of textual data for training, and the vast majority of digital textual data available for training these LLMs is in the English language, and comes from a mix of public and private textual data owned by these institutions. As a result, LLMs have been trained in the language(s) of these globally powerful and financially well-resourced entities.

---

[20] Ma (manuscript-a) suggests this distinction between epistemic and ethical decolonization but in the domain of human ethics of belief in work currently under review.

LLMs have also been trained on other languages, like Chinese – such as Tsinghua University's GLM-130B, Huawei's PanGu LLMs, and Inspur's Yuan 1.0 – but these are typically bilingual models and, in particular, are trained on both English and Chinese language datasets.[21] Currently the largest LLMs are trained on enormous and mostly private English language corpuses (see, e.g., Touvron et al. 2023). Finally, English is also the language in which the majority of the world's academic research, including AI research, is conducted, and in which much of our teaching on AI is done. One of the implications of this fact is that the benchmarks used to test, compare and validate the performance of these large language models are also based in the English language.

Proponents of the linguistic relativity hypothesis have suggested that the particular languages we speak influence which concepts we may have and more generally how we think.[22] While this hypothesis is controversial, numerous empirical studies have suggested, for example, that different languages have different color terms that partition the color spectrum in different ways (see, e.g., Roberson et al. 2005; Lucy 1997; Davies and Corbett 1997; Rosch-Heider 1972; Berlin and Kay 1969; Conklin 1955).

Philosopher Kwasi Wiredu likewise suggests language does shape the way in which we think, and also that, therefore, English and other European languages that have dominated academic discourse and teaching due to colonialism have shaped and in some ways limited the ways in which we think. Wiredu (1985) suggests, for example, that how we understand the concept of "truth" is shaped by the language we speak, and that helpful insights about the nature of truth may be gleaned by English-speaking philosophers by looking to other languages like Akan. Other concepts Wiredu (2002, 58–59) suggests may be linguistically relativistic include *belief*, as well as reality; knowledge; proposition; thought; reason; mind; person; self; objectivity; cause; explanation; chance; nature; space; time; being; existence; object; property; quality; fact; opinion; faith; doubt; certainty; democracy; life; death; god; religion; morality; responsibility; justice; freedom; nature; and more.

Wiredu (2002) hence suggests that we need to engage in what he calls "conceptual decolonization", which he defines as "the elimination from our thought of modes of conceptualization that came to us through colonization and remain in our thinking owing to inertia rather than to our own reflective choice" (Wiredu 2002, 56).

With respect to LLMs, part of this process of linguistic/conceptual decolonization will presumably involve training LLMs in other, especially non-colonial languages; and may also require the design of benchmarks (used to quantify the performance of these LLMs) in a multilingual frame of reference. This linguistic diversification of LLMs and benchmarks will hopefully afford LLMs and society more generally additional and important insights about the world, just as Wiredu suggests consideration of the Akan concept of truth can afford English-speaking philosophers important insights about the nature of truth. And linguistic diversification of LLMs will also of course hopefully portend more inclusivity and access, especially to those who speak non-English languages, and the opportunity for AI research and society as a whole to benefit from their voices and perspectives.

With respect to the *ethical decolonization* of AI too, we suggest, minimally, that alternative ethical systems beyond the ethical systems traditionally considered in the West (e.g., consequentialism, deontology, and virtue ethics) need to continue to be explored as ethical frameworks in the development and deployment of AI. Mhlambi (2020) and others have already begun this important work, for example exploring ubuntu ethics as an alternative ethical framework in which to develop and deploy AI. Indian, Chinese, indigenous and other ethical systems need to be further explored for the same purpose.

---

[21] See, e.g., Jiao et al. (2023); Ren et al. (2023); Wu et al. (2021); and Zeng et al. (2021). And see https://github.com/RUCAIBox/LLMSurvey for a survey of LLMs by Zhao et al. (2023).
[22] See, e.g., Baghramian and Carter (2022) for an accessible introduction to the linguistic relativity hypothesis.

We also suggest, based on Ma (2021; manuscript-a), that we should investigate the effects of the ethical colonization of AI – understood as the domination and subjugation, respectively, of colonial and colonized communities' ethical frameworks in AI – on the moral self-concepts of ethically colonized communities in which ethically colonized AI is deployed. Ethical colonization has often led members of ethically colonized communities to see themselves and to be seen by others as morally inferior, "savage", and uncivilized. The same is also true for epistemically colonized communities who as a result of their epistemic colonization may see themselves and be seen by others as intellectually inferior or less capable of technological advancement, especially in the AI domain.

## 4.2. Epistemic Injustice and AI

Finally, the extensive extant body of philosophical literature on *epistemic injustice* – defined by Miranda Fricker (2007, 1) as "a wrong done as a result of negative identity-prejudicial stereotypes to someone specifically in their capacity as a knower"[23] – is now being applied to AI, where this literature includes content both on AI as perpetrating epistemic injustices as well as AI as constitutionally structured in epistemically unjust ways (where such epistemically unjustly structured AI will often of course perpetrate further epistemic injustices) (see, e.g., Hull 2022; Jacobs & Evers 2023; Pozzi 2023; Sardelli 2022; Stewart et al. 2022; Symons & Alvarado 2022).

Pozzi (2023), for example, suggests that algorithms such as automated Prediction Drug Monitoring Programmes (PDMPs), which are algorithms currently used in the US to predict patients' risk of misusing opioids, may perpetrate *testimonial injustices* against patients. A testimonial injustice is a kind of epistemic injustice in which a speaker receives an unfair deficit of credibility from a hearer owing to prejudice on the hearer's part (Fricker 2007, 154). Pozzi (2023) argues that because PDMP assessments are often used as indicators of trustworthiness and credibility, this leaves open the possibility that patients who are members of marginalized groups will be unjustly discredited given potentially biased PDMP assessments.

Secondly, an algorithm may itself be unjustly epistemically structured in virtue of privileging colonizer perspectives, including colonizer logics, bodies of knowledge, and ethical frameworks as discussed in the previous section (§4.1).[24] Indeed, we take discussion of epistemic injustice and AI to be related to discussion of decolonization of AI in that epistemic colonization is epistemically unjust and leads to epistemically unjust doxastic communities, and achieving epistemic justice will necessarily involve epistemic decolonization. Symons and Alvarado (2022) argue that another way in which AI can be structured in epistemically unjust ways is by being *opaque*, either by design or as a consequence of the sophistication of the algorithm (resulting in, for example, multitudinous and unintuitive features). Opaque or black-box algorithms can hide discriminatory belief-formation, and thwart attempts at recourse for discriminatory beliefs held or decisions made by the algorithm.

More generally, there is much to be explored within the ethics of AI belief respect to epistemic injustice and AI. This is in large part because many of the epistemically and ethically bad and good doxastic states and doxastic practices that should be investigated in an ethics of AI belief will be aptly described as epistemically unjust or epistemically just. Doxastic wronging and doxastic right-doing, for example, if they occur as a result of or in the context of resisting negative identity-prejudicial stereotyping, may be considered epistemically unjust and epistemically just doxastic practices, respectively. Furthermore, given a broader understanding of epistemic injustice as injustice done to agents specifically in their capacities as believers, and epistemic justice as the absence or elimination of epistemic injustice, most of the good and bad doxastic states and practices considered in the ethics of AI belief will be such that they advance either epistemic justice or epistemic injustice overall.

---

[23] McKinnon (2016) points out that although Fricker is the originator of the term "epistemic injustice", Black feminist thinkers and other feminists of color (e.g., Alcoff, 1996; hooks ,1992; Hull, Scott, and Smith, 1982; Ikuenobe, 1998; Collins, 2000; Lorde, 1984; Moraga and Anzaldúa, 1981; Davis, 1981; Spivak, 2003; Alcoff, 2000; Carby, 1982) have been working on issues of epistemic injustice (though not called by that name) for a long time. McKinnon suggests that the fact that the notion of epistemic injustice only secured wide uptake when it was articulated by a white woman is itself a case of epistemic injustice.

[24] On the idea that there may be culturally- or linguistically-specific scientific ontologies see again Wiredu's (2002) notion of "conceptual decolonization", discussed in the previous section (§4.1).

## 5. Conclusion

In sum, we take it that the ethics of AI belief is an exciting new area of research in AI that needs urgently to be explored. The catalogue of specific novel topics proposed for research in §3 was certainly not meant to be an exhaustive list of potential research areas. We do feel, however, that the topics discussed here are important areas for future investigation in AI and have significant implications for social justice, including the identification and elimination of a potential, new kind of discriminatory practice involving doxastic wronging by AI. There are myriad essential epistemic aspects of AI that should not be overlooked in our efforts to develop and deploy ethical and safe AI. We want to stress, furthermore, the importance of interdisciplinary and collaborative work – which will have to begin with working toward establishing common language(s) around key terms like "belief" – between not just ethicists and AI experts but also epistemologists, logicians, philosophers of mind, metaphysicists, and AI experts, as well as experts from other disciplines.

# References

Alcoff, L. M. (1996). The Problem of Speaking for Others. In J. Roof & R. Wiegman (Eds.), *Who Can Speak? Authority and Critical Identity*. University of Illinois Press.

Alcoff, L. M. (2000). On Judging Epistemic Credibility: Is Social Identity Relevant? In N. Zack (Ed.), *Women of Color and Philosophy*. Blackwell Publishers.

Angwin, J., Larson, J., Mattu, S., & Kirchner, L. (2016, May 23). Machine Bias. *ProPublica*. https://www.propublica.org/article/machine-bias-risk-assessments-in-criminal-sentencing

Baghramian, M., & Carter, J. A. (2022). The Linguistic Relativity Hypothesis - Supplement to "Relativism." In *Stanford Encyclopedia of Philosophy* (Spring 202). https://plato.stanford.edu/archives/spr2022/entries/relativism/

Basu, R. (2018a). *Beliefs That Wrong*. University of Southern California.

Basu, R. (2018b). Can Beliefs Wrong? *Philosophical Topics*, *46*(1), 1–18. https://www.jstor.org/stable/26529447

Basu, R. (2019a). The Wrongs of Racist Beliefs. *Philosophical Studies*, *176*, 2497–515. https://doi.org/https://doi.org/10.1007/s11098-018-1137-0

Basu, R. (2019b). What We Epistemically Owe to Each Other. *Philosophical Studies*, *176*, 915–31.

Basu, R. (2019c). Radical Moral Encroachment: The Moral Stakes of Racist Beliefs. *Philosophical Issues*, *29*(1), 9–23. https://doi.org/10.1111/phis.12137

Basu, R. (2021). A Tale of Two Doctrines: Moral Encroachment and Doxastic Wronging. In J. Lackey (Ed.), *Applied Epistemology* (pp. 99–118). Oxford University Press. https://doi.org/10.1093/oso/9780198833659.003.0005

Basu, R. (2023a). Morality of Belief I: How Beliefs Wrong. *Philosophy Compass*, *18*(7).

Basu, R. (2023b). Morality of Belief II: Challenges and Extensions. *Philosophy Compass*, *18*(7).

Basu, R., & Schroeder, M. (2019). Doxastic Wronging. In B. Kim & M. McGrath (Eds.), *Pragmatic Encroachment in Epistemology* (1st ed., pp. 181–205). Routledge.

Beeghly, E. (n.d.). *What's Wrong with Stereotyping*. Oxford University Press.

Beeghly, E. (2015). What is a Stereotype? What is Stereotyping? *Hypatia*, *30*(4), 675–691. https://doi.org/10.1111/hypa.12170

Beeghly, E. (2021). Stereotyping as Discrimination: Why Thoughts Can Be Discriminatory. *Social Epistemology*, 1–17. https://doi.org/10.1080/02691728.2021.1930274

Ben Amor, N., & Benferhat, S. (2005). Graphoid Properties of Qualitative Possibilistic Independence Relations. *International Journal of Uncertainty, Fuzziness and Knowledge-Based Systems*, *13*(01), 59–96. https://doi.org/10.1142/S021848850500331X

Benferhat, S., Leray, P., & Tabia, K. (2020). Belief Graphical Models for Uncertainty Representation and Reasoning. In P. Marquis, O. Papini, & H. Prade (Eds.), *A Guided Tour of Artificial Intelligence Research: Volume II: AI Algorithms* (pp. 209–246). Springer International Publishing. https://doi.org/10.1007/978-3-030-06167-8_8

Benferhat, S., & Smaoui, S. (2007). Hybrid possibilistic networks. *International Journal of Approximate Reasoning*, *44*(3), 224–243. https://doi.org/https://doi.org/10.1016/j.ijar.2006.07.012

Bengio, Y., Lamblin, P., Popovici, D., & Larochelle, H. (2007). Greedy Layer-Wise Training of Deep Networks. *NIPS*.

Bengio, Yoshua. (2009). Learning Deep Architectures for AI. *Foundations and Trends® in Machine Learning*, *2*(1), 1–127. https://doi.org/10.1561/2200000006

Berlin, B., & Kay, P. (1969). *Basic Color Terms: Their Universality and Evolution*. University of California Press.

Bolinger, R. J. (2018). The Rational Impermissibility of Accepting (Some) Racial

Generalizations. *Synthese*, *197*, 2415–2431. https://doi.org/10.1007/s11229-018-1809-5

Bolinger, R. J. (2020). Varieties of Moral Encroachment. *Philosophical Perspectives*, *34*(1), 5–26. https://doi.org/https://doi.org/10.1111/phpe.12124

Bolinger, R. J. (2021). #BelieveWomen and the Ethics of Belief. In M. Schwartzberg & P. Kitcher (Eds.), *NOMOS LXIV: Truth & Evidence*. NYU Press.

Borgelt, C., Gebhardt, J., & Kruse, R. (2000). Possibilistic Graphical Models. In G. Della Riccia, R. Kruse, & H.-J. Lenz (Eds.), *Computational Intelligence in Data Mining* (pp. 51–67). Springer Vienna.

Brownstein, M., & Saul, J. (Eds.). (2016). *Implicit Bias and Philosophy, Volume 2: Moral Responsibility, Structural Injustice, and Ethics*. Oxford University Press. https://doi.org/10.1093/acprof:oso/9780198766179.001.0001

Carby, H. (1982). White Women Listen! Black Feminism and the Boundaries of Sisterhood. In Centre for Contemporary Cultural Studies (Ed.), *The Empire Strikes Back: Race and Racism in the 70s in Britain*. Routledge.

Cave, S. (2020). *The Problem with Intelligence: Its Value-Laden History and the Future of AI*. https://doi.org/10.1145/3375627.3375813

Cave, S., & Dihal, K. (2020). The Whiteness of AI. *Philosophy & Technology*, *33*(4), 685–703. https://doi.org/10.1007/s13347-020-00415-6

Clifford, W. K. (1904). The Ethics of Belief. In L. Stephen & F. Pollock (Eds.), *Lectures and Essays*. Macmillan. https://doi.org/10.1093/0199253722.003.0008

Coates, A., Lee, H., & Ng, A. Y. (2011). An Analysis of Single-Layer Networks in Unsupervised Feature Learning. *Proceedings of the Fourteenth International Conference on Artificial Intelligence and Statistics, PMLR*, 215–223.

Collins, P. H. (2000). *Black Feminist Thought: Knowledge, Consciousness, and the Politics of Empowerment*. Routledge.

Conklin, H. C. (1955). Hanunóo Color Categories. *Southwestern Journal of Anthropology*, *11*(4), 339–344. https://doi.org/10.1086/soutjanth.11.4.3628909

Cozman, F. G. (2000). Credal networks. *Artificial Intelligence*, *120*(2), 199–233. https://doi.org/https://doi.org/10.1016/S0004-3702(00)00029-1

Darwiche, A. (2009). *Modeling and Reasoning with Bayesian Networks*. Cambridge University Press. https://doi.org/DOI: 10.1017/CBO9780511811357

Davies, I. R. L., & Corbett, G. G. (1997). A Cross-Cultural Study of Colour Grouping: Evidence for Weak Linguistic Relativity. *British Journal of Psychology*, *88*(3), 493–517. https://doi.org/https://doi.org/10.1111/j.2044-8295.1997.tb02653.x

Davis, A. Y. (1981). *Women, Race, and Class*. Random House.

de Ville, B. (2013). Decision trees. *WIREs Computational Statistics*, *5*(6), 448–455. https://doi.org/https://doi.org/10.1002/wics.1278

Dennett, D. C. (2010). *Content and Consciousness*. Routledge. https://doi.org/10.4324/9780203092958

Digital Futures Lab, & Research ICT Africa. (2022). *Decolonising AI*. MozFest 2022.

Douglas, H. (2000). Inductive Risk and Values in Science. *Philosophy of Science*, *67*(4), 559–579. http://www.jstor.org/stable/188707

Elliott, K. C., & Richards, T. (Eds.). (2017). *Exploring Inductive Risk: Case Studies of Values in Science*. Oxford University Press. https://doi.org/10.1093/acprof:oso/9780190467715.001.0001

Eubanks, V. (2018). *Automating inequality: how high-tech tools profile, police, and punish the poor* (1st ed.). St. Martin's Press.

Fantl, J., & McGrath, M. (2007). On Pragmatic Encroachment in Epistemology. *Philosophy and Phenomenological Research*, *75*(3), 558–589. https://doi.org/10.1111/j.1933-1592.2007.00093.x

Ferzan, K. K. (2021). #BelieveWomen and the Presumption of Innocence: Clarifying the Questions for Law and Life. In M. Schwartzberg & P. Kitcher (Eds.), *NOMOS LXIV: Truth & Evidence*. NYU Press.

Fodor, J. A. (Jerry A. (1990). *A Theory of Content and Other Essays*. MIT.

Fricker, M. (2007). *Epistemic Injustice: Power and the Ethics of Knowing*. Oxford University Press.

Fritz, J. (2017). Pragmatic Encroachment and Moral Encroachment. *Pacific Philosophical Quarterly*, *98*(S1), 643–661. https://doi.org/10.1111/papq.12203

Gao, J. (2019). Credal pragmatism. *Philosophical Studies*, *176*(6).

Gardiner, G. (2018). Evidentialism and Moral Encroachment. In K. McCain (Ed.), *Believing in Accordance with the Evidence: New Essays on Evidentialism* (pp. 169–195). Springer.

Gelman, A., Carlin, J. B., Stern, H. S., & Rubin, D. B. (2003). Hierarchical models. In R. D. (2003). "Part I. F. of B. D. A. C. . H. models". B. D. A. C. P. pp. 120–. I. 978-1-58488-388-3. Gelman A, Carlin JB, Stern HS (Ed.), *Bayesian Data Analysis*. CRC Press.

Gendler, T. S. (2011). On the Epistemic Costs of Implicit Bias. *Philosophical Studies*, *156*(1), 33–63. http://www.jstor.org/stable/41487720

Grogger, J., Gupta, S., Ivandic, R., & Kirchmaier, T. (2021). Comparing Conventional and Machine-Learning Approaches to Risk Assessment in Domestic Abuse Cases. *Journal of Empirical Legal Studies*, *18*(1), 90–130. https://doi.org/10.1111/JELS.12276

Hadley, R. F. (1991). The Many Uses of 'Belief' in AI. *Minds and Machines*, *1*(1), 55–73. https://doi.org/10.1007/BF00360579

Halpern, J. Y. (2001). Conditional plausibility measures and Bayesian networks. *Journal of Artificial Intelligence Research*, *14*, 359–389.

Heckerman, D. (1995). Tutorial on Learning with Bayesian Networks. In M. I. Jordan (Ed.), *Learning in Graphical Models. Adaptive Computation and Machine Learning* (pp. 301–54). MIT Press.

Hinton, G. (2010). A Practical Guide to Training Restricted Boltzmann Machines. *UTML TR 2010–003*.

Hinton, G. E., & Salakhutdinov, R. R. (2006). Reducing the Dimensionality of Data with Neural Networks. *Science*, *313*(5786), 504–507. https://doi.org/10.1126/science.1127647

Hinton, Geoffrey E. (2009). Deep belief networks. In *Scholarpedia* (p. 4 (5): 5947). https://doi.org/10.4249/scholarpedia.5947

Hinton, Geoffrey E. (1999). Products of Experts. *Proceedings of the Ninth International Conference on Artificial Neural Networks [ICANN 99 Vol 1]*, 1–6.

Hinton, Geoffrey E., Osindero, S., & Teh, Y. W. (2006). A fast learning algorithm for deep belief nets. *Neural Computation*, *18*(7), 1527–1554. https://doi.org/10.1162/neco.2006.18.7.1527

hooks, bell. (1992). *Black Looks: Race and Representation*. South End Press.

Howard, R. A., & Matheson, J. E. (2005). Influence diagrams. *Decision Analysis*, *2*(3), 127–143.

Hull, G. (2022). Dirty Data Labeled Dirt Cheap: Epistemic Injustice in Machine Learning Systems. *SSRN*. https://doi.org/Hull, Gordon, Dirty Data Labeled Dirt Cheap: Epistemic Injustice in Machine Learning Systems (June 15, 2022). Available at SSRN: https://ssrn.com/abstract=4137697 or http://dx.doi.org/10.2139/ssrn.4137697

Hull, G. T., Scott, P. B., & Smith, B. (Eds.). (1982). *All the Women are White, All the Blacks are Men: But Some of Us Are Brave: Black Women's Studies.* Feminist Press at CUNY.

Ikuenobe, P. (1998). A defense of epistemic authoritarianism in traditional African cultures. *Journal of Philosophical Research*, *23*, 417–440.

Jacobs, N., & Evers, J. (2023). Ethical perspectives on femtech: Moving from concerns to capability-sensitive designs. *Bioethics*, *37*(5), 430–439. https://doi.org/https://doi.org/10.1111/bioe.13148

Jensen, F. V. (1996). *An introduction to Bayesian networks* (Vol. 210). University College London Press.

Jiao, F., Ding, B., Luo, T., & Mo, Z. (2023). Panda LLM: Training Data and Evaluation for Open-Sourced Chinese Instruction-Following Large Language Models. *ArXiv Preprint ArXiv:2305.03025*.

Johnson King, Z., & Babic, B. (2020). Moral Obligation and Epistemic Risk. In M. Timmons (Ed.), *Oxford Studies in Normative Ethics, Vol. 10* (pp. 81–105). Oxford University Press.

Johnson King, Z., & Babic, B. (2021). Algorithmic Fairness and Resentment. *Stereotyping and Medical AI*. https://www.youtube.com/watch?v=l3ydHshwzrs

Kahneman, D. (2012). *Thinking, Fast and Slow*. Penguin.

Kelly, D., & Roedder, E. (2008). Racial Cognition and the Ethics of Implicit Bias. *Philosophy Compass*, *3*(3), 522–540. https://doi.org/10.1111/j.1747-9991.2008.00138.x

Kim, B. (2017). Pragmatic Encroachment in Epistemology. *Philosophy Compass*, *12*, 1–14. https://doi.org/10.1111/phc3.12415

Kingsford, C., & Salzberg, S. L. (2008). What are decision trees? *Nature Biotechnology*, *26*(9), 1011–1013. https://doi.org/10.1038/nbt0908-1011

Krishnan, A., Abdilla, A., Moon, A. J., Souza, C. A., Adamson, C., Lach, E. M., Ghazal, F., Fjeld, J., Taylor, J., Havens, J. C., Jayaram, M., Morrow, M., Rizk, N., Quijano, P. R., Çetin, R. B., Chatila, R., Dotan, R., Mhlambi, S., Jordan, S., & Rosenstock, S. (n.d.). *AI Decolonial Manyfesto*. https://manyfesto.ai

Kvanvig, J. L. (2011). Against Pragmatic Encroachment. *Logos and Episteme*, *2*(1), 77–85.

Larochelle, H., & Bengio, Y. (2008). Classification using discriminative restricted Boltzmann machines. *Proceedings of the 25th International Conference on Machine Learning - ICML '08*, 536. https://doi.org/10.1145/1390156.1390224

Lauritzen, S. L., & Spiegelhalter, D. J. (1988). Local computations with probabilities on graphical structures and their application to expert systems. *Journal of the Royal Statistical Society*, *50*, 157–224.

Lee, H., Grosse, R., & Ng, A. Y. (n.d.). *Convolutional Deep Belief Networks for Scalable Unsupervised Learning of Hierarchical Representations*. http://www.cs.toronto.edu/~rgrosse/icml09-cdbn.pdf

Lee, H., Largman, Y., Pham, P., & Ng, A. Y. (n.d.). *Unsupervised feature learning for audio classification using convolutional deep belief networks*. https://ai.stanford.edu/~ang/papers/nips09-AudioConvolutionalDBN.pdf

Leitgeb, H. (2014). The Stability Theory of Belief. *Philosophical Review*, *123*(2), 131–171. https://doi.org/10.1215/00318108-2400575

Lippert-Rasmussen, K. (2013). *Born Free and Equal? A philosophical inquiry into the nature of discrimination*. Oxford University Press.

Loar, B. (1981). *Mind and Meaning*. Cambridge University Press.

Lorde, A. (1984). *Sister Outsider: Essays and Speeches*. Crossing Press.

Lucy, J. A. (1997). The Linguistics of "Color." In C. L. Hardin & L. Maffi (Eds.), *Color Categories in Thought and Language* (pp. 320–346). Cambridge University Press. https://doi.org/DOI: 10.1017/CBO9780511519819.015

Ma, W. (n.d.). *Doxastic Self-Confidence, Epistemic Injustice, & Epistemic Colonization*.

Ma, W. (2021). *A Pragmatist Ethics of Belief* [King's College London]. https://kclpure.kcl.ac.uk/portal/en/persons/winnie-ma(50b82b74-d335-4148-af83-a814ef2524ab)/theses.html

Ma, W. (2022a). Bounded Emotionality and Our Doxastic Norms. *Inquiry*. https://doi.org/10.1080/0020174X.2022.2124540

Ma, W. (2022b). Profiling in Public Health. In S. Venkatapuram & A. Broadbent (Eds.), *The Routledge Handbook of the Philosophy of Public Health* (pp. 161–175). Routledge.

Ma, W. (2023). *Doxastic Wronging (by Failing to Believe and Failing to Believe In) & Doxastic Right-Doing*.

McHugh, Conor. 2012. "The Truth Norm of Belief." *Pacific Philosophical Quarterly* 93: 8–30.

McKinnon, R. (2016). Epistemic Injustice. *Philosophy Compass*, *11*, 437–446.

Mhlambi, S. (2020). From Rationality to Relationality: Ubuntu as an Ethical and Human Rights Framework for Artificial Intelligence Governance. *Carr Center Discussion Paper Series*. https://carrcenter.hks.harvard.edu/files/cchr/files/ccdp_2020-009_sabelo_b.pdf

Mitova, V. (2022). White Ignorance Undermines Internalism about Epistemic Blame. *Royal Institute of Philosophy Lecture*.

Moraga, C., & Anzaldúa, G. (Eds.). (1981). *This Bridge Called My Back: Writings by Radical Women of Color*. Persephone Press.

Moss, S. (2018). Moral Encroachment. *Proceedings of the Aristotelian Society*, *118*(2), 177–205. https://doi.org/10.1093/arisoc/aoy007

Mugg, J. (2013). What are the Cognitive Costs of Racism? A Reply to Gendler. *Philosophical Studies*, *166*(2), 217–229. https://doi.org/10.1007/s11098-012-0036-z

Nelson, M. T. (2010). We Have No Positive Epistemic Duties. *Mind*, *119*(473), 83–102. http://www.jstor.org/stable/40865208

Nikolaiev, D. (2023, March). Behind the Millions: Estimating the Scale of Large Language Models. *Medium*. https://towardsdatascience.com/behind-the-millions-estimating-the-scale-of-large-language-models-97bd7287fb6b

Pearl, J. (1986). Fusion, Propagation, and Structuring in Belief Networks. *Artificial Intelligence*, *29*(3), 241–288. https://doi.org/https://doi.org/10.1016/0004-3702(86)90072-X

Pearl, J. (1988). *Probabilistic Reasoning in Intelligent Systems: Networks of Plausible Inference* (2nd ed.). Morgan Kaufmann.

Pearl, J. (1990). Reasoning with belief functions: An analysis of compatibility. *International Journal of Approximate Reasoning*, *4*(5), 363–389. https://doi.org/https://doi.org/10.1016/0888-613X(90)90013-R

Pearl, J. (2005). Influence Diagrams—Historical and Personal Perspectives. *Decision Analysis*, *2*(4), 232–234. https://doi.org/10.1287/deca.1050.0055

Peels, R. (2017). *Responsible Belief: A Theory in Ethics and Epistemology*. Oxford University Press. https://doi.org/10.1093/acprof:oso/9780190608118.001.0001

Perlis, D. (2000). The Role(s) of Belief in AI. In J. Minker (Ed.), *Logic-Based Artificial Intelligence* (pp. 361–374). Springer US. https://doi.org/10.1007/978-1-4615-1567-8_16

Pozzi, G. (2023). Testimonial Injustice in Medical Machine Learning. *Journal of Medical Ethics*, jme-2022-108630. https://doi.org/10.1136/jme-2022-108630

Puddifoot, K. (2017). Dissolving the Epistemic/Ethical Dilemma Over Implicit Bias. *Philosophical Explorations*, *20*(sup1), 73–93. https://doi.org/10.1080/13869795.2017.1287295

Puddifoot, K. (2019). Stereotyping Patients. *Journal of Social Philosophy*, *50*(1), 69–90. https://doi.org/10.1111/josp.12269

Qian, H., Marinescu, R., Gray, A., Bhattacharjya, D., Barahona, F., Gao, T., Riegel, R., & Sahu, P. (2021). Logical Credal Networks. *ArXiv Preprint*.

Raiffa, H. (1968). *Decision analysis: introductory lectures on choices under uncertainty*. Addison-Wesley.

Ren, X., Zhou, P., Meng, X., Huang, X., Wang, Y., Wang, W., Li, P., Zhang, X., Podolskiy, A., & Arshinov, G. (2023). PanGu-{\Sigma}: Towards Trillion Parameter Language Model with Sparse Heterogeneous Computing. *ArXiv Preprint ArXiv:2303.10845*.

Research ICT Africa. (2022). *Decolonising AI: Ethics and the Rule of Law*.

Rinard, S. (2019). Equal Treatment for Belief. *Philosophical Studies*, *176*(7), 1923–1950.

https://doi.org/10.1007/s11098-018-1104-9

Roberson, D., Davidoff, J., Davies, I. R. L., & Shapiro, L. R. (2005). Color Categories: Evidence for the Cultural Relativity Hypothesis. *Cognitive Psychology*, *50*(4), 378–411.

Robinson, A., Myhill, A., Wire, J., Roberts, J., & Tilley, N. (2016). *Risk-led policing of domestic abuse and the DASH risk model*. https://pure.southwales.ac.uk/en/publications/risk-led-policing-of-domestic-abuse-and-the-dash-risk-model

Rosch-Heider, E. R. R. (1972). Universals in Color Naming and Memory. *Journal of Experimental Psychology*, *93*(1), 10–20. https://doi.org/info:doi/

Russell, S. J., & Norvig, P. (2003). *Artificial Intelligence: A Modern Approach* (2nd ed.). Prentice Hall.

Sackett, D. L., Rosenberg, W. M. C., Gray, J. A. M., Haynes, R. B., & Richardson, W. S. (1996). Evidence Based Medicine: What It Is and What it Isn't. *BMJ*, *312*, 71–72. https://doi.org/10.1136/bmj.312.7023.71

Sardelli, M. (2022). Epistemic Injustice in the Age of AI. *Aporia*, *22*(XXII).

Schofield, M. J., & Mishra, G. D. (2003). Validity of Self-Report Screening Scale for Elder Abuse: Women's Health Australia Study. *The Gerontologist*, *43*(1), 110–120. https://doi.org/10.1093/GERONT/43.1.110

Schwartz, R., Dodge, J., Smith, N. A., & Etzioni, O. (2020). Green AI. *Communications of the ACM*, *63*(12), 54–63.

Schwitzgebel, E. (2021). Belief. In *Stanford Encyclopedia of Philosophy*. http://plato.stanford.edu/entries/belief/

Shachter, R. D. (1986). Evaluating influence diagrams. *Operations Research*, *34*(6), 871–882.

Shenoy, P. P. (1989). A valuation-based language for expert systems. *International Journal of Approximate Reasoning*, *3*(5), 383–411. https://doi.org/https://doi.org/10.1016/0888-613X(89)90009-1

Smithies, D. (2015). Ideal Rationality and Logical Omniscience. *Synthese*, *192*(9), 2769–2793. http://www.jstor.org/stable/24704815

Smolensky, P. (1986). Information Processing in Dynamical Systems: Foundations of Harmony Theory. In D. E. Rumelhart & J. L. McLelland (Eds.), *Parallel Distributed Processing: Explorations in the Microstructure of Cognition, Volume 1: Foundations* (pp. 194–281). MIT Press.

Spivak, G. C. (2003). Can the Subaltern Speak? *Die Philosophin*, *14*(27), 42–58.

Stanley, J. (2005). *Knowledge and Practical Interests*. Oxford University Press. https://doi.org/10.1093/0199288038.001.0001

Steel, D. (2010). Epistemic Values and the Argument from Inductive Risk. *Philosophy of Science*, *77*(1), 14–34. https://doi.org/DOI: 10.1086/650206

Stewart, H., Cichocki, E., & McLeod, C. (2022). A Perfect Storm for Epistemic Injustice: Algorithmic Targeting and Sorting on Social Media. *Feminist Philosophy Quarterly*, *8*(3/4).

Strubell, E., Ganesh, A., & McCallum, A. (2019). Energy and Policy Considerations for Deep Learning in NLP. *ArXiv Preprint ArXiv:1906.02243*.

Symons, J., & Alvarado, R. (2022). Epistemic Injustice and Data Science Technologies. *Synthese*, *200*(2), 87. https://doi.org/10.1007/s11229-022-03631-z

Thaler, R. H. (2015). *Misbehaving: The Making of Behavioral Economics* (1st ed.). W.W. Norton & Company.

Touvron, H., Lavril, T., Izacard, G., Martinet, X., Lachaux, M.-A., Lacroix, T., Rozière, B., Goyal, N., Hambro, E., & Azhar, F. (2023). Llama: Open and Efficient Foundation Language Models. *ArXiv Preprint ArXiv:2302.13971*.

Vaithianathan, R., Benavides-Prado, D., Dalton, E., Chouldechova, A., & Putnam-Hornstein, E. (2021). Using a Machine Learning Tool to Support High-Stakes Decisions in Child Protection. *AI Magazine*, *42*(1), 53–60. https://ojs.aaai.org/aimagazine/index.php/aimagazine/article/view/7482

Van Leeuwen, N., & Lombrozo, T. (2023). The Puzzle of Belief. *Cognitive Science*, *47*(2), e13245. https://doi.org/https://doi.org/10.1111/cogs.13245

Vredenburgh, K. (2023). AI and bureaucratic discretion. *Inquiry*, 1–30. https://doi.org/10.1080/0020174X.2023.2261468

Wiredu, K. (1985). The Concept of Truth in the Akan Language. In P. O. Bodunrin (Ed.), *Philosophy in Africa: Trends and Perspectives*. University of Ife Press.

Wiredu, K. (2002). Conceptual Decolonization as an Imperative in Contemporary African Philosophy: Some Personal Reflections. *Rue Descartes*, *36*(2), 53–64. https://doi.org/10.3917/rdes.036.0053

Worsnip, A. (2020). Can Pragmatists Be Moderate? *Philosophy and Phenomenological Research*, 1–28. https://doi.org/10.1111/phpr.12673

Wu, S., Zhao, X., Yu, T., Zhang, R., Shen, C., Liu, H., Li, F., Zhu, H., Luo, J., & Xu, L. (2021). Yuan 1.0: Large-Scale Pre-Trained Language Model in Zero-Shot and Few-Shot Learning. *ArXiv Preprint ArXiv:2110.04725*.

Xu, H., & Smets, P. (1994). Evidential reasoning with conditional belief functions. In *Uncertainty Proceedings 1994* (pp. 598–605). Elsevier.

Yaghlane, B. Ben, & Mellouli, K. (2008). Inference in directed evidential networks based on the transferable belief model. *International Journal of Approximate Reasoning*, *48*(2), 399–418. https://doi.org/https://doi.org/10.1016/j.ijar.2008.01.002

Zeng, W., Ren, X., Su, T., Wang, H., Liao, Y., Wang, Z., Jiang, X., Yang, Z., Wang, K., & Zhang, X. (2021). PanGu-$\alpha $: Large-scale Autoregressive Pretrained Chinese Language Models with Auto-parallel Computation. *ArXiv Preprint ArXiv:2104.12369*.

Zhao, W. X., Zhou, K., Li, J., Tang, T., Wang, X., Hou, Y., Min, Y., Zhang, B., Zhang, J., & Dong, Z. (2023). A Survey of Large Language Models. *ArXiv Preprint ArXiv:2303.18223*.

Zimmerman, A. Z. (2018). *Belief: A Pragmatic Picture*. Oxford University Press. https://doi.org/10.1093/oso/9780198809517.001.0001